\documentclass{article}\usepackage[]{graphicx}\usepackage[]{color}
\pdfoutput=1
\makeatletter
\def\maxwidth{ %
  \ifdim\Gin@nat@width>\linewidth
    \linewidth
  \else
    \Gin@nat@width
  \fi
}
\makeatother

\definecolor{fgcolor}{rgb}{0.345, 0.345, 0.345}
\newcommand{\hlkwb}[1]{\textcolor[rgb]{0.69,0.353,0.396}{#1}}%

\usepackage{framed}
\makeatletter
 {\par\unskip\endMakeFramed%
 \at@end@of@kframe}
\makeatother

\definecolor{shadecolor}{rgb}{.97, .97, .97}
\definecolor{messagecolor}{rgb}{0, 0, 0}
\definecolor{warningcolor}{rgb}{1, 0, 1}
\definecolor{errorcolor}{rgb}{1, 0, 0}
\usepackage{alltt}

\usepackage[english]{babel}
\usepackage[utf8x]{inputenc}
\usepackage[T1]{fontenc}
\usepackage{multirow}
\usepackage{cite}
\usepackage{arxiv}
\usepackage{hyperref}       % hyperlinks
\usepackage{url}            % simple URL typesetting
\usepackage{booktabs}       % professional-quality tables
\usepackage{amsfonts}       % blackboard math symbols
\usepackage{nicefrac}       % compact symbols for 1/2, etc.
\usepackage{microtype}      % microtypography
\usepackage{lipsum}

\usepackage{amsmath}
\usepackage{graphicx}
\usepackage[colorinlistoftodos]{todonotes}
\usepackage{float}
\usepackage{array}

\newcommand\independent{\protect\mathpalette{\protect\independenT}{\perp}}
\def\independenT#1#2{\mathrel{\setbox0\hbox{$#1#2$}%
\copy0\kern-\wd0\mkern4mu\box0}}

\title{The covariate-adjusted residual estimator and its use in both randomized trials and observational settings}
\author{
Stephen A.~Lauer \\
Department of Epidemiology\\
Bloomberg School of Public Health\\
Johns Hopkins University \\
Baltimore, MD 21205 \\
\texttt{slauer5@jh.edu} \\
\AND
Nicholas G.~Reich \\
Department of Biostatistics\\
University of Massachusetts \\
Amherst, MA 01003\\
\texttt{nick@schoolph.umass.edu} \\
\AND
Laura B.~Balzer \\
Department of Biostatistics\\
University of Massachusetts \\
Amherst, MA 01003\\
\texttt{lbalzer@schoolph.umass.edu} \\
}
\IfFileExists{upquote.sty}{\usepackage{upquote}}{}
\begin{document}

\maketitle

\begin{abstract}
We often seek to estimate the causal effect of an exposure on a particular outcome in both randomized and observational settings.
One such estimation method is the covariate-adjusted residuals estimator, which was designed for individually or cluster randomized trials.
In this manuscript, we study the properties of this estimator and develop a new estimator that utilizes both covariate adjustment and inverse probability weighting.
We support our theoretical results with a simulation study and an application in an infectious disease setting.
The covariate-adjusted residuals estimator is an efficient and unbiased estimator of the average treatment effect in randomized trials; however, it is not guaranteed to be unbiased in observational studies.
Our novel estimator, the covariate-adjusted residuals estimator with inverse probability weighting, is unbiased in randomized and observational settings, under a reasonable set of assumptions.
Furthermore, when these assumptions hold, it provides efficiency gains over inverse probability weighting in observational studies.
The covariate-adjusted residuals estimator is valid for use in randomized trials, but should not be used in observational studies.
The covariate-adjusted residuals estimator with inverse probability weighting provides an efficient alternative for use in randomized and observational settings.
\end{abstract}

Estimating the effect of an exposure on a population has a long history.
In 1855, John Snow compared the mortality rates of households in London by the company that supplied their water to locate the source of a cholera epidemic.\cite{Snow1855}
In 1881, Louis Pasteur inoculated 50 sheep with anthrax, 25 of whom had been vaccinated; the vaccinated sheep survived as the unvaccinated died, providing evidence that his anthrax vaccine was effective.\cite{Merrill2010}
In 1948, Austin Bradford Hill conducted the first modern randomized clinical trial, to evaluate a treatment for pulmonary tuberculosis,\cite{Hill1948, Yoshioka1998} and later formulated guidelines for researchers and practitioners to transition from thinking in terms of statistical association to those of causation.\cite{Hill1965}
Since then, there has been a proliferation of methods to determine the exposure effects in randomized trials and observational studies.\cite{Fisher1932, Horvitz1952, Cochran1957, Cox1982, Rosenbaum1983, Robins1986, VanderLaan2006, Tsiatis2008, Hayes2009, Moore2009, Rosenblum2010, Rose2011, Balzer2016}

One such method is the covariate-adjusted residuals estimator (CARE), which was formulated to estimate the effect of an exposure on an outcome of interest in individually randomized or cluster randomized trials.\cite{Gail1996, Bennett2002, Hayes2009}
To implement CARE, researchers first predict the outcome using baseline covariates that influence the outcome, while leaving out the exposure.
Then they find the average residual for each group; this error can be the difference between or the ratio of the observed values versus predicted values (i.e., the prediction error).
The CARE estimate of the exposure effect is the discrepancy between the average residuals in each group.
In a randomized trial setting, \cite{Gail1996} used parametric regression models to predict the outcome as a function of baseline covariates, and demonstrated that while maintaining confidence interval coverage, CARE could increase the statistical power over an unadjusted estimator, which is the average difference (ratio) in outcomes between randomized groups.
\cite{Bennett2002} showed that CARE made accurate estimates of the effect size in cluster randomized trials, even when there were small numbers of clusters and moderate imbalances in the distribution of predictive covariates.
Most recently, CARE was used to investigate the effect of a Universal Test-and-Treat intervention on HIV incidence in a large cluster randomized trial in Zambia and South Africa.\cite{hayes_impact_2019}
To the best of our knowledge, the properties of CARE in a randomized trial have not been evaluated non-parametrically.

In observational settings, CARE is commonly used in ecology under the name `residual index',\cite{Jakob1996, Kotiaho2001, Blackwell2002, ELKIN2005, Sanchez-Chardi2007, Danielson-Francois2009} although it has received some criticism.\cite{Green2001,Garcia-Berthou2001}
In the ecological field of allometry, researchers have used the residual index to estimate the association of an exposure on the body mass of an organism, often in observational settings rather than randomized trials.
While there are domain-specific questions about whether ordinary least squares linear regression is being used appropriately in allometry,\cite{Green2001} others have questioned the statistical validity of the residual index under any circumstances.
In particular, \cite{Garcia-Berthou2001} stated that ``even if the assumptions of the linear model hold for the original variables, they will not hold for the residuals'' and thus ``the `residual index' should never be used for statistical analyses of condition or any other variable''.
To the best of our knowledge, there has been no statistical theory presented to date to support the continued use of the residual index, and thus CARE, in an observational setting.

In this manuscript, we provide new non-parametric theory that shows CARE is an unbiased estimator of the effect of a binary exposure in randomized trials, but a biased estimator in most observational settings of interest. We also provide the conditions in which CARE is a consistent estimator and asymptotically normal.
Our work supplements and generalizes existing parametric results from \cite{Gail1996} for randomized trials and establishes new theory for observational studies.
We compare CARE to existing estimators and introduce a novel estimator for use in both randomized and observational settings; CARE-IPW, our new estimator, combines CARE with methods using inverse probability weighting.\cite{Horvitz1952, Robins2000, Rosenbaum1983} We prove CARE-IPW is an unbiased estimator in both randomized trials and observational settings, and provide the conditions for consistency and asymptotic normality.
We support these theoretical results with a simulation study.

As an illustration, we apply these methods to real data to estimate the effect of bednets on childhood mortality in a cluster randomized trial in Ghana as originally published by \cite{Binka1996} and discussed in \cite{Bennett2002, Hayes2009}.
In this trial, the Kassena-Nankana region of Ghana was divided into 96 clusters, 48 of which were randomly selected to receive impregnated bednets in June 1993.
From July 1993 to June 1995, children aged 6-59 months were surveilled until they died (the primary outcome), they migrated out of the study area, they turned 60 months of age, or the end of the follow-up period was reached.
The clusters had 138 to 439 children each, with an average of 274.4 children.
The data includes age in months at time of enrollment, sex, outcome, years of follow-up, and the cluster-level exposure assignment for each child.
To improve the precision of their analysis, \cite{Binka1996} used CARE to adjust for the imbalanced age distributions between the exposure levels.
The authors found that bednets were associated with an all-cause child mortality rate ratio of 0.83 (95\% CI: 0.69-1.00).
\cite{Hayes2009} repeated this analysis using both age and sex as covariates and compared the estimated rate ratio from CARE to that of an unadjusted analysis.
Both the CARE rate ratio (RR: 0.844, 95\% CI: 0.713, 0.999) and rate difference (RD: -4.26, 95\% CI: -8.76, 0.23) estimates indicated a stronger association than that found by the unadjusted estimator (RR: 0.859, 95\% CI: 0.721, 1.023; RD: -3.95, 95\% CI: none provided).\cite{Hayes2009}
We will use this case study as an example when describing the causal framework and statistical theory and as our real data application.

\section{Causal Roadmap}
\label{sec:causal}

Did bednets reduce childhood mortality in Ghana?
This is a common structure for a causal scientific question: how would a change in an exposure (e.g. bednets) change an outcome (e.g. childhood mortality).
As a result, answering causal questions require a different approach than descriptive or associative questions.
For example, a descriptive analysis may provide point and uncertainty estimates for the childhood mortality in clusters that actually received bednets and in clusters that did not receive bednets.
If we were interested in predicting childhood mortality, we would want to know whether including a covariate for bednets added value to predictions that may use other information (such as age and sex), regardless of whether that relationship was causal or associative.
Answering the causal question requires a deeper understanding of the system that generates the exposure and the outcome, as well as the influence of additional covariates.

Several conceptual and analytic frameworks exist and can guide our answering of causally motivated questions.\cite{Neyman1923, Rubin1974, Robins1986, Rubin1990, Holland1986, Spirtes1993, Pearl2009, Robins2009, Richardson2013swig,Petersen2014, balzer_tutorial_2016,  Hernan2016}.
Here, we review the Causal Roadmap of \cite{Petersen2014} and use the bednet example for illustration.
The key steps of the Causal Roadmap are representing knowledge of the data generating process (represented by a \textit{causal model}); specifying the quantity that answers the scientific question (the \textit{causal parameter}); evaluating the assumptions required to link the causal quantity to a well-defined function of the observed data distribution (the \textit{statistical parameter} which may or may not be \textit{identifiable}); and finally obtaining a point estimate and inference of the statistical parameter.

\begin{figure}
\includegraphics[width=0.49\textwidth]{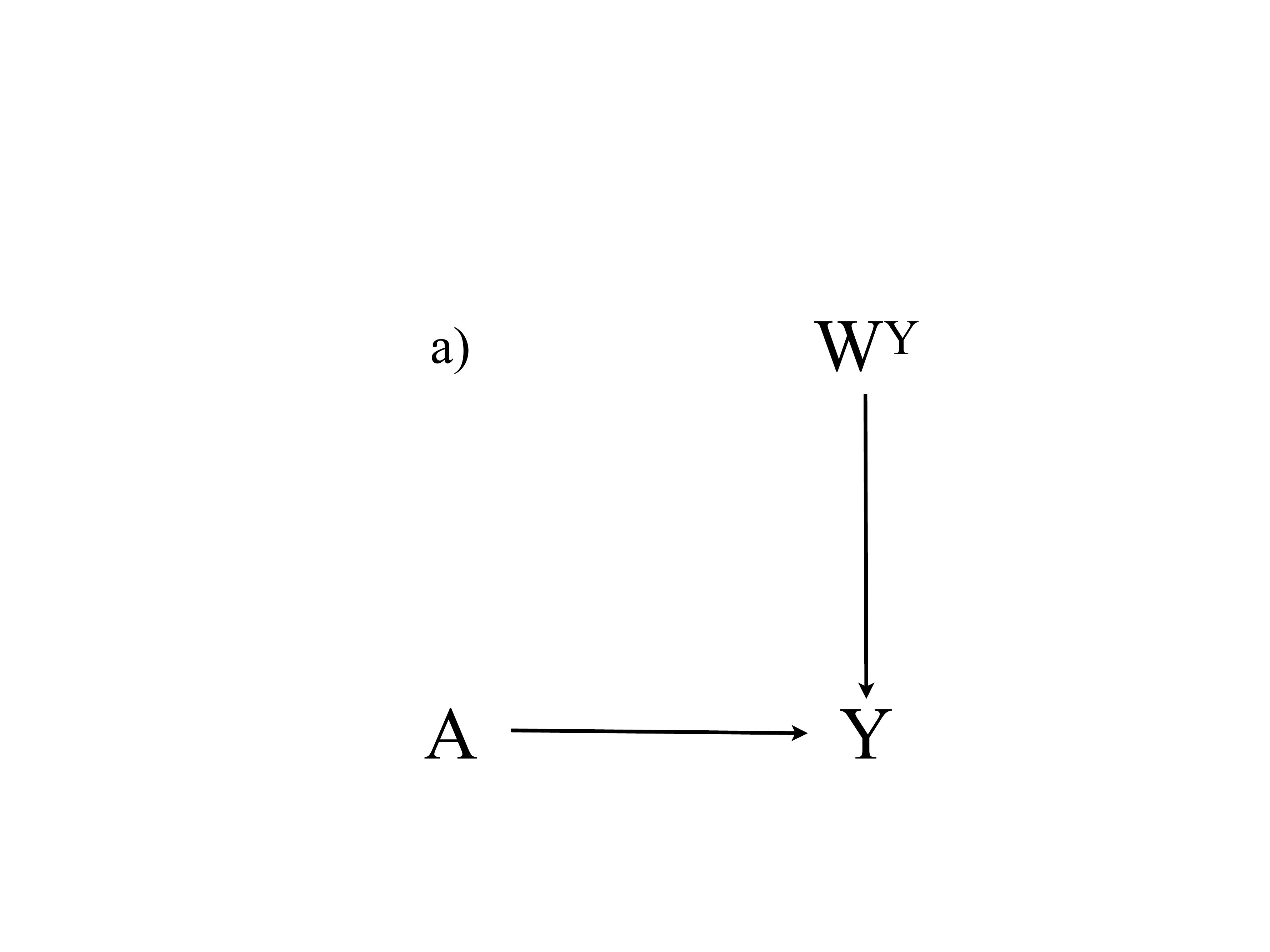}
\includegraphics[width=0.49\textwidth]{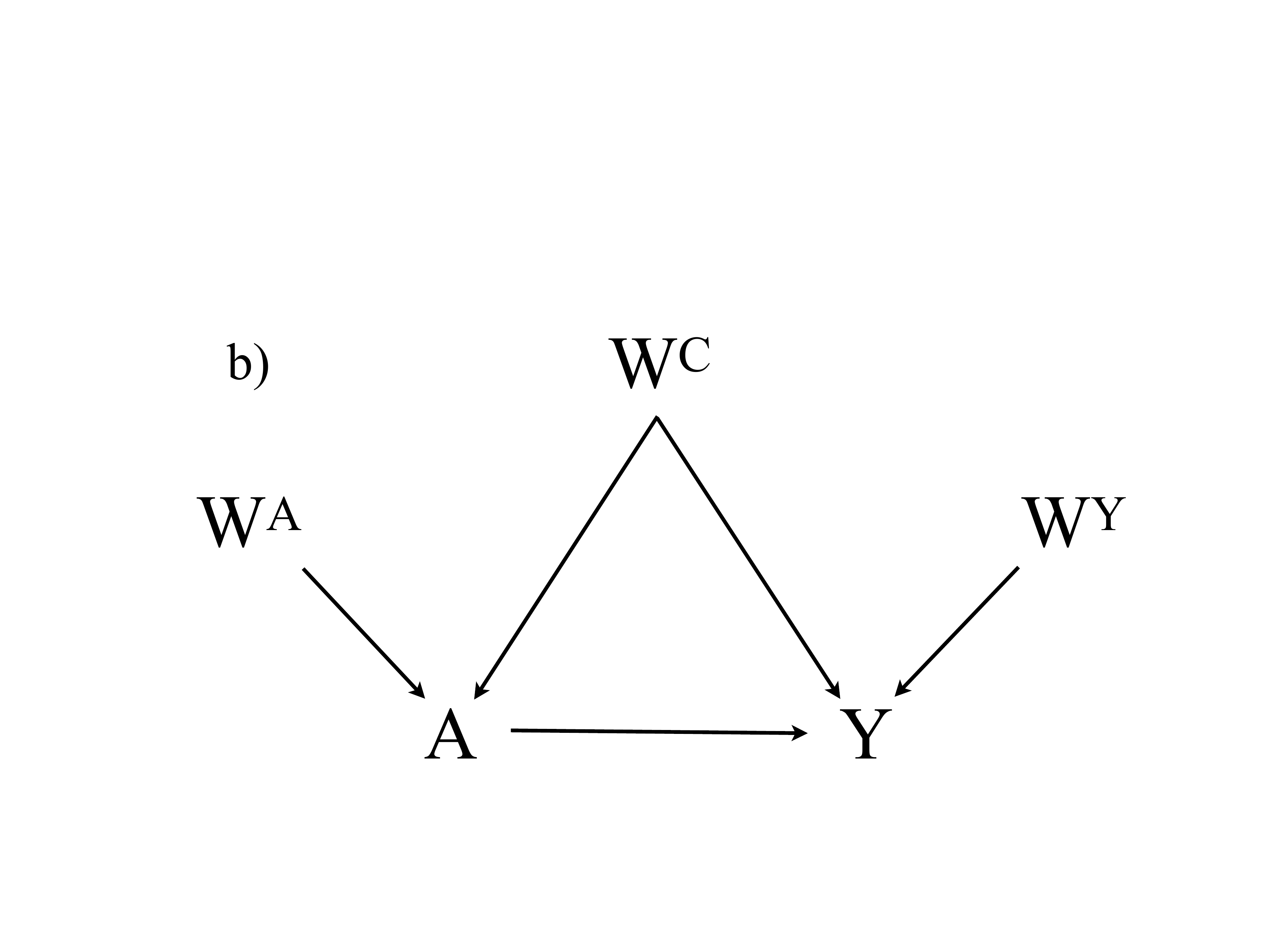}
\caption{
Causal diagrams for randomized trials \textbf{(a)} and observational studies \textbf{(b)}.
These diagrams give a visual representation of the relationships between the variables in a causal model.
Arrows are drawn from a potential cause to an effect.
In a completely randomized trial setting, the exposure $A$ is generated independently of all other variables and the outcome $Y$ may be influenced by both the exposure $A$ and a set of baseline covariates $W^Y$.
In an observational setting, the exposure $A$ is no longer randomized, but instead is influenced by baseline covariates.
Some of these covariates $W^C$ also influence $Y$, thus confounding the relationship between the exposure $A$ and the outcome $Y$.
Other covariates $W^A$ only influence the exposure $A$ and not the outcome $Y$; as before some covariates $W^Y$ only influence the outcome $Y$ and not the exposure $A$. (For simplicity, other unmeasured sources of variation are omitted; see the Supplemental Material for a complete graph).
}
\label{fig:dag}
\end{figure}

A causal model is a structural framework for expressing the relationships between variables in a given setting.\cite{Pearl1988, Pearl2009, Pearl2010, Goldberger1972, Duncan1975}
A causal model can be expressed graphically as a diagram, where variables are connected by edges (arrows) that originate at a potential cause and terminate at the effect.
Figure \ref{fig:dag}a is a diagram representing a randomized trial, like that of our case study, where $A$ is a binary exposure ($A=1$ if the cluster received the bednet intervention, $A=0$ if the cluster did not) and $W^Y$ is the set of baseline covariates (the average age and percentage of children who are female for each cluster) that may influence the outcome $Y$ (childhood mortality).
There are no edges pointing to the exposure $A$ because the randomization procedure makes the allocation of exposure independent of all other covariates.
(A diagram including unmeasured variables is depicted in the Supplementary Materials.)
For this experimental setting, we assume that this causal model describes the data generating process for each cluster and that clusters are causally independent (i.e. the outcome for one cluster is only influenced by that cluster's exposure and baseline covariates and independent of other clusters' exposures, baseline covariates, and outcomes.)

In an observational setting (as portrayed in Figure \ref{fig:dag}b), the allocation to the exposure $A$ is not randomized and is potentially influenced by the baseline covariates.
In addition to the covariates $W^Y$ that influence the outcome $Y$, but not the exposure $A$, there are two new subsets of covariates.
One subset of covariates $W^A$ only influence the exposure $A$, but not the outcome $Y$.
The other subset are confounding covariates $W^C$ that influence both the exposure $A$ and the outcome $Y$ and thus obscure the isolation of the causal effect of interest.
As a running example, suppose bednets are distributed to clusters by the determination of local health officials instead of at random.
In this scenario, consider a new baseline covariate: mosquito abundance.
Places with greater mosquito abundance prior to the intervention may be higher risk for infectious diseases and future childhood mortality and public health officials would want to concentrate their efforts in these areas.
Thus mosquito abundance is a confounding covariate, as it is a common cause of both the outcome and the exposure.

With the causal model specified, we can translate our scientific question into a causal parameter.
We assume that the relationships within the causal model are autonomous, meaning that changing one relationship does not change the other relationships, although changes may result in different effects downstream.\cite{Pearl2009,Pearl2010}
Therefore, we could modify the way in which the exposure is generated and see resulting changes in the outcome.
Specifically, we could intervene to give impregnated bednets to a cluster (i.e. set $A=1$)
and generate the counterfactual (potential) outcome $Y(1)$ for that cluster.
Likewise, we could intervene to put the same cluster in the unexposed group (i.e. set $A=0)$
and generate the counterfactual (potential) outcome $Y(0)$ for that cluster.
With these counterfactual outcomes, we translate this scientific question into a well-defined causal quantity, specifically the average treatment effect (ATE):
\begin{align}
ATE = \mathbb{E}[Y(1)-Y(0)]. \label{eqn:ate}
\end{align}
The ATE is the expected difference in the average childhood mortality rate if all of the clusters in our target population received impregnated bednets ($Y(1)$) and if none of the clusters received impregnated bednets ($Y(0)$).
We cannot directly estimate this parameter because we only observe the outcomes $Y$ corresponding to the actual exposures $A$ and not both counterfactual outcomes.
Thus, we need to outline the conditions and assumptions necessary to identify the causal parameter as a statistical parameter of the observed data distribution.

In an observational setting, suppose our observed data consist of confounding covariates $W^C$, the exposure indicator $A$, and the outcome $Y$; we denote the observed data as $O=(W^C, A, Y)$ and assume we have $n$ independent, identically distributed (i.i.d.) copies from some distribution $\mathbb{P}$.
In this setting, we use the difference in conditional expectations between the exposed and unexposed, adjusted for and averaged across the measured confounding covariates, as the statistical parameter:\begin{align}
\Psi^{obs}(\mathbb{P}) &= \mathbb{E}_{W^C} \big[ \mathbb{E}(Y \mid A=1,W^C) - \mathbb{E}(Y \mid A=0,W^C) \big]. \label{eqn:gcomp}
\end{align}
This statistical parameter $\Psi^{obs}(\mathbb{P})$ is known as the ``G-computation identifiability result'', which identifies the ATE under two assumptions.\cite{Robins1986}
First, there must be no unmeasured confounding between the exposure and the outcome: $Y(a) \independent A \mid W^C$.
Second, the `positivity assumption', which states that each possible strata of measured confounding covariates has a non-zero probability of being in each exposure group ($\mathbb{P}(A=a | W^C=w^C)>0, \forall w^C \in \mathbb{P}(W^C=w^C)>0$), must hold.\cite{Petersen2010}
In our example, the statistical parameter $\Psi^{obs}(\mathbb{P})$ is the expected difference in the childhood mortality rate between clusters with and without bednets adjusting for common causes.

\begin{figure}[hbt!]
\centering
\includegraphics[width=\linewidth]{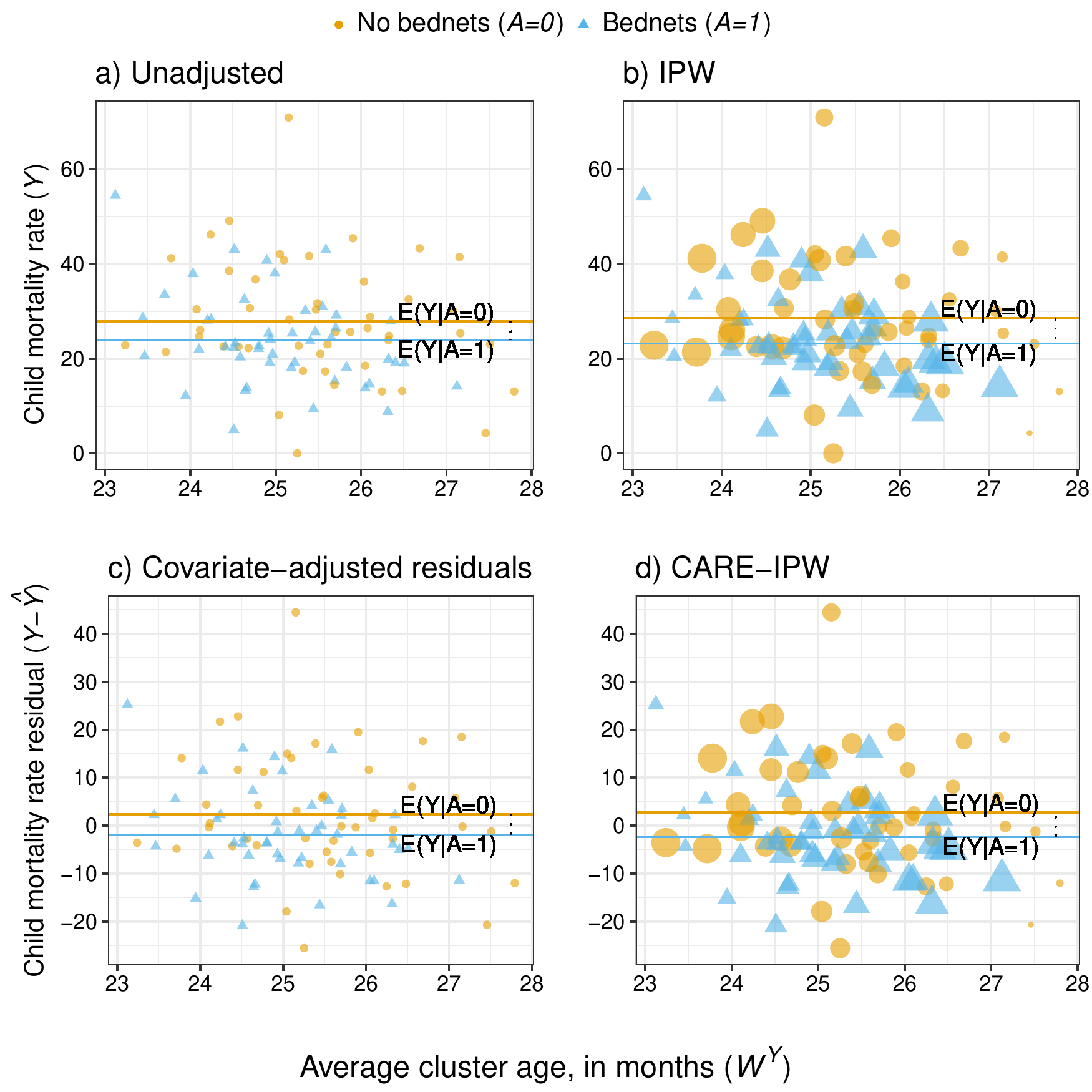}
\caption{
Estimates of the effect of bednets on childhood mortality rate using different methods; data obtained from \cite{Hayes2009}.
In all plots, the clusters are arranged by the average age of the children in the cluster (x-axis);  blue triangles denote clusters assigned to receive bednets ($A=1$; intervention), and orange circles denote clusters not assigned to receive bednets ($A=0$; control).
Despite randomization, there were imbalances in baseline covariates predictive of the outcome ($W^Y$) between arms.
\textbf{(a)} The unadjusted estimate is the difference in the average child mortality rate ($Y$), per thousand follow-up years, between intervention clusters (dashed blue line) and control clusters (solid orange line).
\textbf{(b)} The inverse probability weighting (IPW) estimator gives greater weight to intervention clusters with higher average ages and control clusters with lower average ages, as indicated by the size of the triangles and circles.
The point estimate from IPW is the difference in the average of the weighted outcomes between groups.
\textbf{(c)} The covariate-adjusted residuals estimator (CARE) uses predictions ($\hat{Y}$) of the child mortality rates based on Poisson regression with average age and proportion female as covariates.
The point estimate from CARE is the difference in the average residuals between intervention clusters and control clusters.
\textbf{(d)} The covariate-adjusted residuals estimator with inverse probability weighting (CARE--IPW) combines the weights from the IPW estimator with the residuals from CARE.
Specifically, the point estimate from CARE--IPW is the difference in the average of the weighted residuals between groups.
}
\label{fig:example}
\end{figure}

In a randomized trial, the process for allocating units to the exposure is truly random (i.e. a coin flip). Therefore, the assumptions of no unmeasured confounding and positivity are satisfied naturally by the study design.
Therefore, in this setting, we can identify the ATE with the target statistical parameter $\Psi^{RCT}(\mathbb{P})$, the difference in the expectation of the outcome between exposure groups:
\begin{align}
\Psi^{RCT}(\mathbb{P}) = \mathbb{E}(Y|A=1) - \mathbb{E}(Y|A=0) \label{eqn:rct}.
\end{align}
The statistical parameter $\Psi^{RCT}(\mathbb{P})$ can be consistently estimated using the difference in the average outcome between the exposure groups, also known as the `unadjusted estimator':\cite{Neyman1923}
\begin{align}
\hat{\Psi}^{unadj}(\hat{\mathbb{P}}) &= \frac{1}{n_1} \sum_{i \forall A_i=1} Y_i - \frac{1}{n_0} \sum_{i \forall A_i=0} Y_i = \hat{\mathbb{E}}(Y|A=1) - \hat{\mathbb{E}}(Y|A=0), \label{eqn:unadj}
\end{align}
where $\hat{\mathbb{P}}$ is the empirical distribution; $n_a$ is the number of units ($i=1,\dots,n=n_0+n_1$) in exposure level $A=a$, and $\hat{\mathbb{E}}(Y|A=a)$ is an estimate of the expected outcome in exposure group $A=a$.
In the bednet cluster randomized trial, the unadjusted estimate is the difference between the average mortality rate for clusters assigned to receive bednets and the average mortality rate for clusters not assigned to receive bednets, as illustrated in Figure \ref{fig:example}a.

In observational settings, the exposure allocation is no longer random, and we have to account for common causes of the exposure $A$ and outcome $Y$.
If $W^C$ captures all those common causes (i.e. there is no unmeasured confounding) and there is sufficient variability in the exposure within possible strata of $W^C$ (i.e. positivity holds), then the inverse probability weighting (IPW) estimator can be used to estimate the statistical parameter $\Psi^{obs}(\mathbb{P})$:\cite{Horvitz1952,Robins2000}
\begin{align}
\hat{\Psi}^{IPW}(\hat{\mathbb{P}}) &= \frac{1}{n} \sum_{i=1}^n \left(\frac{\mathbb{I}(A_i=1)}{\hat{\mathbb{P}}(A=1 \mid W^C_i)} - \frac{\mathbb{I}(A_i=0)}{1-\hat{\mathbb{P}}(A=1 \mid W^C_i)} \right) Y_i, \label{eqn:ipw}
\end{align}
which controls for the confounding covariates through estimates of the conditional probability of receiving the exposure, called `propensity scores' $\hat{\mathbb{P}}(A=1 \mid W^C)$.\cite{Rosenbaum1983}
Intuitively, the IPW estimator up-weights exposure-covariate combinations that are rare, relative to a randomized trial, and down-weights exposure-covariate combinations that are more common, relative to a randomized trial.
If the propensity scores are consistent for the true conditional probability of exposure given common causes $\mathbb{P}(A=1 \mid W^C)$, the IPW estimator is consistent for the statistical parameter $\Psi^{obs}(\mathbb{P})$.
Notably, the propensity scores do not need to account for other covariates that only influence the exposure $W^A$.

IPW can also be used in randomized trials, and in that setting estimating the known propensity score can lead to efficiency gains over the unadjusted estimator.\cite{vanderLaan2003, Shen2014, Balzer2016}
In the bednet cluster randomized example, we used a logistic regression with independent variables for average age and the proportion of children that are female to estimate the probability of bednet assignment for each cluster $\hat{\mathbb{P}}(A=1 \mid W^Y)$.
Despite randomization, we found that the estimated propensity scores ranged from 0.22 to 0.72 and were higher for clusters with lower average age than for clusters with higher average age.
In the study, younger clusters were, by chance, more likely to receive the bednet intervention than older clusters.
When obtaining a point estimate by taking the average difference in weighted outcomes, intervention clusters with higher average ages were assigned greater weight, as are control clusters with lower average ages (Figure \ref{fig:example}b).

\section{The covariate-adjusted residuals estimator (CARE)}
\label{sec:care}

The covariate-adjusted residuals estimator (CARE) was proposed as a method to estimate the ATE in randomized trials.\cite{Gail1996, Bennett2002, Hayes2009}
First, the outcome $Y$ is  predicted using only the baseline covariates $W^Y$ and not the exposure $A$, giving us predicted values $\hat{\mathbb{E}}(Y \mid W^Y)$.
Next, residuals are derived as the difference between the observed outcome and the predicted one.
Finally, the difference in the average residuals between exposure groups provides a point estimate:
\begin{align}
\hat{\Psi}^{CARE}(\hat{\mathbb{P}}) &= \underbrace{\frac{1}{n_1} \sum_{i \forall A_i=1} \left[Y_i - \hat{\mathbb{E}}(Y \mid W^Y_i) \right]}_{\text{Average residual for exposed}} \ - \ \underbrace{ \frac{1}{n_0} \sum_{i \forall A_i=0} \left [Y_i - \hat{\mathbb{E}}(Y \mid W^Y_i) \right]}_{\text{Average residual for unexposed}} \label{eqn:care} \\
&= \frac{1}{n}\sum_{i=1}^n \left( \frac{\mathbb{I}(A_i=1)}{\hat{\mathbb{P}}(A=1)} - \frac{\mathbb{I}(A_i=0)}{\hat{\mathbb{P}}(A=0)} \right) \left( Y_i - \hat{\mathbb{E}}(Y \mid W^Y_i) \right), \label{eqn:care3}
\end{align}
where the number of units at each exposure level is equal to the total number of units times the empirical probability of exposure $n_a = n \times \hat{\mathbb{P}}(A=a)$.
To obtain the predictions of the outcome in the absence of the exposure $\hat{\mathbb{E}}(Y \mid W^Y)$, \cite{Hayes2009} recommend using Poisson regression for event rates, logistic regression for binary outcomes, and linear regression for continuous outcomes.

In the bednet cluster randomized example, we used Poisson regression to estimate the child mortality rate for each cluster with independent variables for average age and the proportion of children that are female, but not for bednet assignment.
The point estimate from CARE is then the difference between the average residual for intervention clusters and the average residual for control clusters (Figure \ref{fig:example}c)

\textbf{Theorem 2.1}
\textit{In a randomized trial, the covariate adjusted residuals estimator (CARE) is an unbiased estimator of the target statistical parameter $\Psi^{RCT}(\mathbb{P}) = \mathbb{E}(Y|A=1) - \mathbb{E}(Y|A=0)$ and thus the average treatment effect (ATE) because the identifiability assumptions hold by design.}

The proof is given in Appendix \ref{app:care-rct}. Briefly, consider the following estimating function of the observed data $O=(W^Y,A,Y)$ and parameter $\psi$: \cite{Rose2011, Kennedy2017}
\begin{align}
D(O; \psi) =\left( \frac{\mathbb{I}(A=1)}{\mathbb{P}(A=1)} - \frac{\mathbb{I}(A=0)}{\mathbb{P}(A=0)} \right) \left( Y -\mathbb{E}(Y \mid W^Y) \right) - \psi
\label{eqn:CAREee}
\end{align}
$D$ is an unbiased estimating function for $\Psi^{RCT}(\mathbb{P})=\psi^{RCT}$ in that when $\psi=\psi^{RCT}$ its expectation is zero: $\mathbb{E}[D(O; \psi^{RCT}]=0$ (proof in Appendix \ref{app:care-rct}).
The corresponding estimating equation is given by
\begin{align*}
0 = \frac{1}{n} \sum_{i=1}^n D(O_i; \psi)
\end{align*}
and we obtain a point estimate from CARE by solving this estimating equation. In other words,
$\hat{\psi}^{CARE}$ is the solution satisfying $0=1/n \sum_{i=1}^n \hat{D}(O_i; \hat{\psi}^{CARE})$, as shown in Equation~\ref{eqn:care3}.

CARE requires estimation of both the marginal probability of the exposure $\mathbb{P}(A=1)$ and conditional expectation of the outcome, given the covariates $\mathbb{E}(Y \mid W^Y)$.
Since the exposure mechanism is always consistently estimated in randomized trials, CARE will be consistent for $\Psi^{RCT}(\mathbb{P})$ in a trial setting.
Under regularity conditions,\cite{Rose2011, Kennedy2017} the Central Limit Theorem applies, and CARE is asymptotically normal with variance well-approximated by the sample variance of $\hat{D}(O;\hat{\psi}^{CARE})$ divided by sample size $n$.
This simple variance estimator can be used for construction of Wald-Type 95\% confidence intervals and testing the null hypothesis.

When accounting for predictive covariates $W^Y$ that are imbalanced by chance between the two randomized groups, we expect CARE to provide efficiency gains over the unadjusted estimator. \cite{Fisher1932, Cochran1957, Cox1982, Tsiatis2008, Moore2009, Rosenblum2010, Balzer2016} When the predicted values of the outcome $\hat{\mathbb{E}}(Y \mid W^Y)$ are a constant value (e.g. zero or the mean of all observations $\hat{\mathbb{E}}(Y)$), then CARE is equivalent to the unadjusted estimator (proof in Appendix \ref{app:care-rct}).
Thus, the unadjusted estimator could be considered a special case of CARE.

\textbf{Theorem 2.2}
\textit{In most observational settings of interest, the covariate adjusted residuals estimator (CARE) is a biased estimator of the target statistical parameter $\Psi^{obs}(\mathbb{P})=\mathbb{E}_{W^C}\big[ \mathbb{E}(Y|A=1, W^C) - \mathbb{E}(Y|A=0,W^C) \big]$.}

The proof is given in Appendix~\ref{app:care-obs}.
Briefly, the expectation of the CARE estimating function (Equation~\ref{eqn:CAREee}) for $\Psi^{obs}(\mathbb{P}) = \psi^{obs}$ is
\begin{align*}
\mathbb{E} & \left[ D(O; \psi^{obs}) \right] = \psi^{RCT} - \mathbb{E} \left[ \left( \frac{\mathbb{I}(A=1)}{\mathbb{P}(A=1)} - \frac{\mathbb{I}(A=0)}{\mathbb{P}(A=0)}\right) \mathbb{E}(Y \mid W^C) \right] - \psi^{obs}.
\end{align*}
For this expectation to be zero, we would need the residual difference between the ``unadjusted'' estimand $\psi^{RCT}$ and the ``adjusted'' estimand $\psi^{obs}$ to be captured by the average difference in the weighted predictions in the absence of the exposure $\mathbb{E}(Y|W^C)$.
There is no reason to believe this would generally be the case.
Instead, we can only expect this to hold under the strong null, where $\psi^{obs}=0$ and $\mathbb{E}(Y \mid A, W^C) = \mathbb{E}(Y \mid W^C)$.
When the null is false (i.e. there is an exposure effect), there might also be scenarios where we get some bias cancellation, but this cannot be proven under a non-parametric statistical model.
Thus, CARE is only guaranteed to be a consistent estimator of $\Psi^{obs}(\mathbb{P})$ when the null is true and conditional mean outcome $\mathbb{E}(Y \mid A, W^C) = \mathbb{E}(Y \mid W^C)$ is consistently estimated.
Since we do not know \textit{a priori} whether or not the null hypothesis is true (which is presumably why we are trying to estimate the exposure effect), we do not recommend that CARE be used in observational settings.

\section{Improving upon CARE with inverse probability weighting}
\label{sec:care-ipw}

As previously discussed, the inverse probability weighting (IPW) estimator controls for measured confounders by up-weighting observations that have a rare exposure-covariate combination (relative to a randomized trial) and down-weighting those with a common exposure-covariate combination (again relative to a randomized trial) using the estimated propensity scores $\hat{\mathbb{P}}(A=1 \mid W^C)$.
This suggests that we may improve CARE (Equation~\ref{eqn:care3}) for use in observational settings by replacing the empirical probabilities of exposure $\hat{\mathbb{P}}(A=1)$ with estimated propensity scores $\hat{\mathbb{P}}(A=1 \mid W^C)$.
To our knowledge, such an estimator has not been previously proposed or explored, and thus we name it ``covariate-adjusted residuals estimator with inverse probability weighting'' (CARE--IPW):
\begin{align}
\hat{\Psi}^{CARE-IPW}(\mathbb{\hat{P}}) &= \frac{1}{n}\sum_{i=1}^n  \left( \frac{ \mathbb{I}(A_i=1)}{\hat{\mathbb{P}}(A=1 \mid W^C_i)} - \frac{ \mathbb{I}(A_i=0)}{\hat{\mathbb{P}}(A=0 \mid W^C_i)} \right) \left( Y_i - \hat{\mathbb{E}}(Y\mid W^C_i) \right). \label{eqn:care-ipw}
\end{align}
The CARE--IPW estimate is the difference in the weighted average of the residuals for the intervention group and for the control group.

The CARE--IPW estimator can also be applied to randomized trials by replacing the confounding covariates $W^C$, which are not present in randomized trials, with the covariates that affect the outcome $W^Y$.
In the bednet cluster randomized trial example, we used Poisson regression to estimate the child mortality rate for each cluster $\mathbb{E}(Y \mid W^Y)$, as in the CARE estimator, and a logistic regression to estimate the propensity scores for bednet assignment $\mathbb{P}(A=1 \mid W^Y)$, as in the IPW estimator (Figure \ref{fig:example}d).

\textbf{Theorem 3.1}
\textit{In an observational setting, the covariate-adjusted residuals estimator with inverse probability weighting (CARE--IPW) is an unbiased estimator of the target statistical parameter $\Psi^{obs}(\mathbb{P}) = \mathbb{E}_{W^C}\big[ \mathbb{E}(Y|A=1, W^C) - \mathbb{E}(Y|A=0,W^C) \big]$.
In randomized trials where the identifiability assumptions hold by design and $\Psi^{obs}(\mathbb{P})=\Psi^{RCT}(\mathbb{P})$, CARE--IPW is an unbiased estimator of the average treatment effect (ATE).}

The proof is given in Appendix \ref{app:care-ipw}.
Briefly, consider the following estimating function of the observed data $O=(W^C,A,Y)$ and parameter $\psi$:
\begin{align}
D^*(O; \psi) =\left( \frac{\mathbb{I}(A=1)}{\mathbb{P}(A=1 \mid W^C)} - \frac{\mathbb{I}(A=0)}{\mathbb{P}(A=0 \mid W^C)} \right) \left( Y -\mathbb{E}(Y \mid W^C) \right) - \psi
\label{eqn:CAREIPWee}
\end{align}
$D^*$ is an unbiased estimating function for $\Psi^{obs}(\mathbb{P})=\psi^{obs}$ in that when $\psi=\psi^{obs}$ its expectation is zero: $\mathbb{E}[D^*(O; \psi^{obs}]=0$ (proof in Appendix \ref{app:care-ipw}).
The corresponding estimating equation is given by
\begin{align*}
0 = \frac{1}{n} \sum_{i=1}^n D^*(O_i; \psi)
\end{align*}
We obtain a point estimate from CARE--IPW by solving this estimating equation. In other words,
$\hat{\psi}^{CARE-IPW}$ is the solution satisfying $1/n \sum_{i=1}^n  \hat{D}^*(O_i; \hat{\psi}^{CARE-IPW})$, as shown in Equation~\ref{eqn:care-ipw}.

CARE--IPW requires estimation of both the propensity score $\mathbb{P}(A=1 \mid W^C)$ and conditional expectation of the outcome, given the covariates $\mathbb{E}(Y \mid W^C)$.
CARE--IPW is consistent when the propensity score $\mathbb{P}(A=1 \mid W^C)$ is consistently estimated, or when the null is true and conditional mean outcome $\mathbb{E}(Y \mid A, W^C) = \mathbb{E}(Y \mid W^C)$ is consistently estimated.
Under regularity conditions,\cite{Rose2011, Kennedy2017} the Central Limit Theorem applies, and CARE--IPW is asymptotically normal with variance well-approximated by the sample variance of $\hat{D}^*(O;\hat{\psi}^{CARE-IPW})$ divided by sample size $n$.
By predicting the outcome with the covariates $\hat{\mathbb{E}}(Y \mid W^C)$, we expect CARE-IPW to provide efficiency gains over the IPW estimator.

Each of the estimators described in this paper can be characterized as special cases of CARE--IPW.
CARE--IPW reduces to CARE when the propensity scores are estimated with the empirical probability of exposure.
CARE--IPW reduces to IPW when the predicted values of the outcome are all zero.
CARE--IPW reduces to the unadjusted estimator when both of the above conditions are met.

\section{Simulation Study}

In this section, we use a simulation to compare the performance of the unadjusted estimator, IPW, CARE, and CARE--IPW in a randomized trial as well as an observational setting.
We consider a synthetic data-generating process with a binary exposure and a binary outcome.
In the randomized setting, three covariates affect the outcome.
In the observational setting, the relationship between the exposure and the outcome is confounded by two covariates.
In both settings, there is no unmeasured confounding and positivity holds by design; estimates can, therefore, be interpreted causally.

We compare the performance of the estimators using bias, Monte Carlo standard error, average standard error estimate, confidence interval coverage, power, and type I error.
Let $\hat{\psi}_{s}$ denote the point estimate in simulation $s$, $s=1,\dots,S$.
Bias is the average difference between the point estimate and the statistical parameter $\frac{1}{S}\sum_{s=1}^S \hat{\psi}_{s}-\psi$, where $\psi$ is $\psi^{obs}$ in the observational setting and $\psi^{rct}$ in the randomized setting.
Monte Carlo standard error is the standard error of the point estimates across simulations $\sqrt{Var(\hat{\psi}_{1:S})}$.
The average standard error estimate is the mean of the estimated standard errors across simulations $\frac{1}{S}\sum_{s=1}^S \sqrt{\nu_{s}}$, where $\nu_s$ is the influence curve-based estimate of the variance in simulation $s$.
Confidence interval coverage is the proportion of 95\% confidence intervals that covered the statistical parameter $\psi$ across all simulations.
Power is the proportion of simulations that the estimator rejected the null hypothesis of no exposure effect when there was an exposure effect.
Type I error is the proportion of simulations that the estimator rejected the null hypothesis of no exposure effect when the null hypothesis was true.

All simulations were run using R version 3.4.3.\cite{R2018}
Simulations were run in parallel on 15 cores on a remote server.
To maintain reproducibility and to make sure that the same samples were drawn for each scenario (with and without an effect in randomized and observational settings) across simulations, we set a seed for the random number generator for each simulation based on the simulation number.
The code used for this project can be found at https://doi.org/10.5281/zenodo.3517241.

\subsection{Setup}

To study the finite sample properties of CARE and CARE--IPW relative to those of the unadjusted and IPW estimators, we designed a synthetic simulation with binary exposures and outcomes.

Consider an experiment with 96 units.
For each unit in the sample, we generate four independent baseline covariates: $W1 \sim Normal(0,1)$, $W2 \sim Normal(0,1)$, $W3 \sim Uniform(0,1)$, and $W4 \sim Bernoulli(0.5)$.
We simulate a randomized trial where the exposure $A$ is assigned with probability 0.5 as well as an observational setting where the exposure is assigned with a probability given by $logit^{-1}[1 -0.75 \cdot W1 -2 \cdot W4 + 0.5 \cdot W2]$.
Each unit's counterfactual outcomes, $Y(1)$ and $Y(0)$, are generated as
\begin{align*}
Y(A) &= Bernoulli\left(logit^{-1}[ -0.25 + 0.5 \cdot W1 - 1 \cdot W3 + 2 \cdot W4 - 1.25 \cdot A - 0.5 \cdot A \cdot W3] \right)
\end{align*}
by deterministically setting the exposure to $A=1$ and $A=0$, respectively.
The average treatment effect is calculated by taking the mean difference in the counterfactual outcomes for a population of 100,000 units.
We also simulate a scenario under the null hypothesis of no exposure effect by setting the counterfactual outcome with the exposure $Y(1)$ equal to the counterfactual outcome without the exposure $Y(0)$.

We implement the unadjusted estimator as the difference in average outcomes between exposed and unexposed units (Equation~\ref{eqn:unadj}).
When estimating the propensity score, required for IPW and CARE--IPW, we use a logistic regression with main terms for $W1$ and $W4$, which are the confounders in the observational setting.
For the outcome prediction, which is required for CARE and CARE--IPW, we use a logistic regression with main terms for $W1$, $W3$, and $W4$, which corresponds to the correctly specified regression under the null.

\subsection{Results}

Table \ref{table-sim} provides a comparison of the performance of the estimators over $S$=5,000 repetitions of the simulation.
When there is an effect, the intervention $A$ led to a -28.1\% average reduction in the outcome.

All estimators are unbiased in the randomized trial setting.
The 95\% confidence interval coverage for each algorithm is close to or above the nominal level.
Improvements in Monte Carlo standard error, average standard error, and statistical power over both the unadjusted and the IPW estimators are achieved by both CARE and CARE--IPW.

% latex table generated in R 3.5.3 by xtable 1.8-3 package
% Thu Oct 24 10:18:22 2019
\begin{table}[ht]
\centering
\caption{Results for the estimators for the simulation by trial type and exposure effect. The covariate-adjusted residuals estimator (CARE) uses a logistic regression with $W1$, $W3$, and $W4$ to predict the outcome. The inverse probability of weighting (IPW) estimator uses a logistic regression with $W1$ and $W4$ to estimate the propensity scores. CARE with inverse probability weighting (CARE--IPW) the same regression as CARE to predict the outcome and the same regression as IPW to estimate the propensity scores.} 
\label{table-sim}
\begin{tabular}{lllrrrb{1.25cm}b{1.2cm}}
  \hline
Trial & Exposure & Estimator & Bias & MC SE & Average SE & 95\% CI coverage & Power/ Type I error \\ 
  \hline
\hline \multirow{4}{*}{RCT} & \multirow{4}{*}{Effect} & CARE--IPW & 0.003 & 0.092 & 0.092 & 94.5\% & 85.4\% \\ 
   &  & CARE & 0.008 & 0.090 & 0.090 & 94.4\% & 85.3\% \\ 
   &  & IPW & -0.001 & 0.094 & 0.148 & 99.7\% & 46.2\% \\ 
   &  & Unadj & -0.002 & 0.101 & 0.101 & 94.3\% & 78.1\% \\ 
  \hline \multirow{4}{*}{RCT} & \multirow{4}{*}{Null} & CARE--IPW & 0.000 & 0.093 & 0.090 & 94.1\% & 5.9\% \\ 
   &  & CARE & 0.000 & 0.091 & 0.089 & 94.4\% & 5.6\% \\ 
   &  & IPW & 0.000 & 0.095 & 0.167 & 99.9\% & 0.1\% \\ 
   &  & Unadj & -0.000 & 0.104 & 0.103 & 94.4\% & 5.6\% \\ 
  \hline \multirow{4}{*}{Obs} & \multirow{4}{*}{Effect} & CARE--IPW & 0.000 & 0.115 & 0.115 & 94.5\% & 71\% \\ 
   &  & CARE & 0.062 & 0.082 & 0.081 & 87.4\% & 75.6\% \\ 
   &  & IPW & -0.005 & 0.126 & 0.164 & 98.7\% & 44.6\% \\ 
   &  & Unadj & -0.197 & 0.088 & 0.089 & 41.7\% & 100\% \\ 
  \hline \multirow{4}{*}{Obs} & \multirow{4}{*}{Null} & CARE--IPW & -0.004 & 0.107 & 0.102 & 94.1\% & 5.9\% \\ 
   &  & CARE & -0.003 & 0.079 & 0.087 & 96.7\% & 3.3\% \\ 
   &  & IPW & -0.005 & 0.124 & 0.197 & 99.6\% & 0.4\% \\ 
   &  & Unadj & -0.219 & 0.100 & 0.099 & 39.7\% & 60.3\% \\ 
   \hline
\end{tabular}
\end{table}

When there is an effect in the observational setting, the unadjusted estimator is markedly biased with low confidence interval coverage: 41.7\%.
By adjusting for confounders when predicting the outcome, CARE reduces but does not eliminate bias and achieves confidence interval coverage of 87.4\%, still much less than the nominal level.
Through consistent estimation of the propensity score and thereby control for the confounders, both the IPW estimator and CARE--IPW are unbiased and achieve nominal to conservative confidence interval coverage.
CARE--IPW is more efficient and achieves higher statistical powerful than the IPW estimator: 71.0\% vs. 44.6\%, respectively.

When there is no effect in an observational setting, the unadjusted estimator is again biased with low confidence interval coverage at 39.7\%.
Both CARE, and CARE--IPW are unbiased with nominal to conservative Type I error control and greater precision than the IPW estimator.
We note that under the null, CARE is expected to be consistent if the outcome is correctly predicted, which it was here.

Altogether this simulation confirms the theoretical properties described in Sections \ref{sec:care}-\ref{sec:care-ipw}.

\section{Case study}

In this section, we first reproduce the findings of \cite{Hayes2009}, who compared CARE to the unadjusted estimator in the cluster randomized trial to estimate the impact of impregnated bednets on child mortality in northern Ghana.\cite{Bennett2002, Binka1996}
Then we apply the IPW estimator and CARE--IPW on the same data and discuss the results.

\subsection{Setup}

In the original analysis, the researchers estimated the unadjusted and covariate-adjusted mortality rates for the exposed and unexposed groups and compared them using the \textit{t}-test.
For the unadjusted estimator, the observed mortality rate (\textit{i.e.} the number of deaths per thousand follow-up years) was calculated for each cluster.
The unadjusted estimate of the ATE was equal to the difference in the average observed mortality rate between randomized arms (Figure \ref{fig:example}a).

In the covariate-adjusted analysis, the researchers used a Poisson regression for mortality rate on the individual-level data using age and sex as covariates, but not the cluster intervention assignment.
From this regression, they predicted the mortality rate per follow-up year for each child, which they then aggregated into cluster-level predicted mortality rates per thousand follow-up years.
The researchers found the residuals by taking the difference between the observed and predicted mortality rates for each cluster.
The CARE estimate of the ATE was equal to the difference in the average of the residuals between randomized arms (Figure \ref{fig:example}c).
Hayes and Moulton used a $t$-test to generate confidence intervals and conduct hypothesis testing.

We reproduce this analysis and extend it to include IPW and CARE--IPW (Figure \ref{fig:example}b,d).
While our point estimates of the ATE for CARE and the unadjusted estimator are identical to those in Hayes and Moulton, we estimate the variance using influence curve-based methods (Sections \ref{sec:care}-\ref{sec:care-ipw}), which yield slightly different confidence intervals and \textit{p}-values.
The IPW and CARE--IPW estimators require propensity scores, which we estimate with a main terms logistic regression for the exposure at the cluster-level using average age in months and percent of children who are female as covariates.
For CARE--IPW, we use the same predicted values of the outcome from the individual-level regression as used for CARE.

\subsection{Results}

\begin{figure}
\centering
\includegraphics[width=1\linewidth]{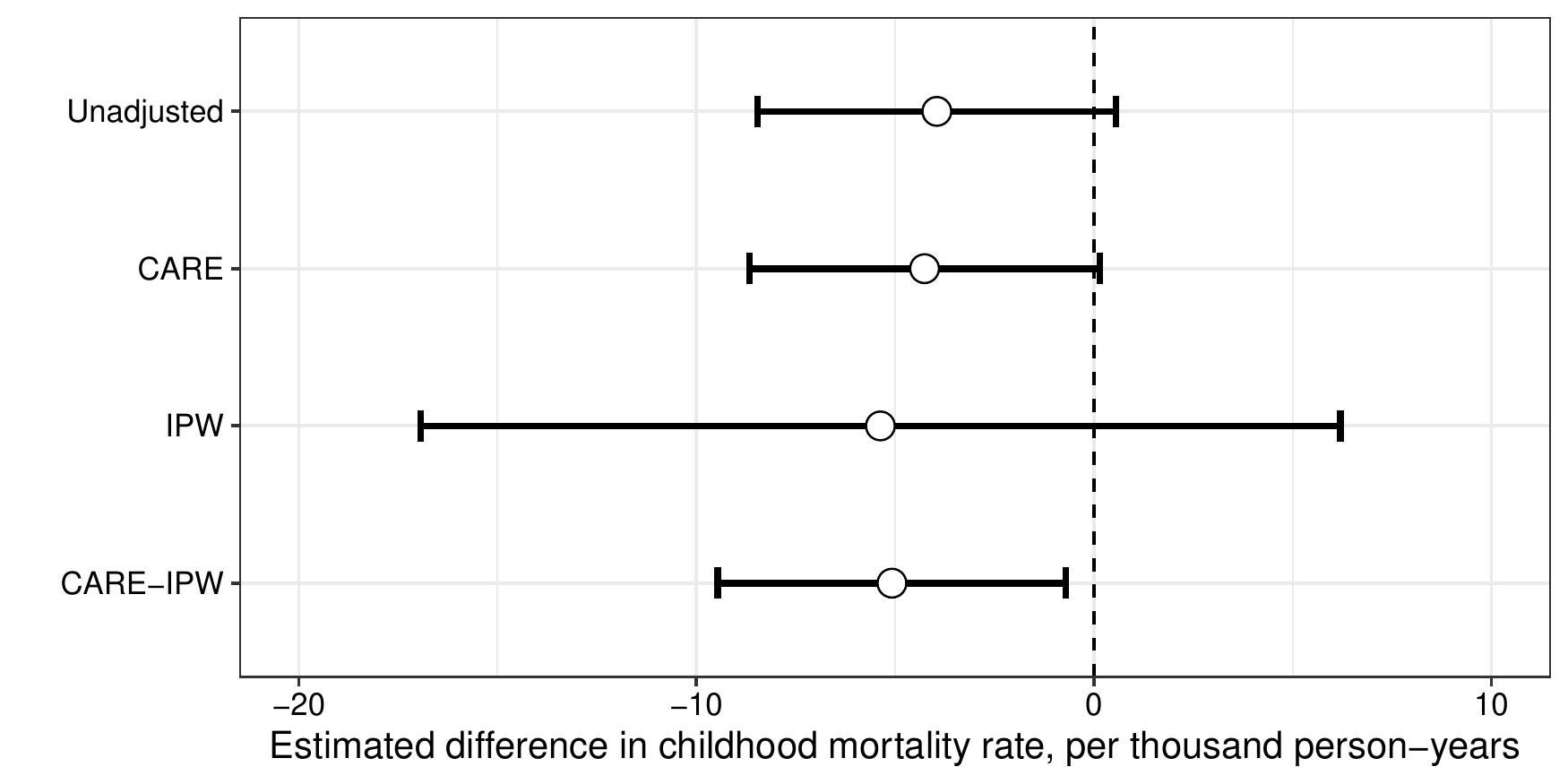}
\caption[The estimates and 95\% confidence intervals for the effect of allocating bednets on childhood mortality rate per thousand follow-up years.]{The estimates and 95\% confidence intervals for the effect of allocating bednets on childhood mortality rate per thousand follow-up years; data obtained from \cite{Hayes2009}.
The four algorithms are the unadjusted estimator, the covariate-adjusted residuals estimator (CARE), the inverse probability weighting estimator (IPW), and CARE with inverse probability weighting (CARE--IPW).
While all estimates indicate that bednets cause a reduction in childhood mortality, the CARE and CARE--IPW estimates are more precise than those of the unadjusted and IPW estimators.}\label{fig:application-plot}
\end{figure}

The IPW and CARE--IPW estimates of the exposure effect are larger than the estimates from the unadjusted estimator or CARE (Figure \ref{fig:application-plot}).
As in the original analysis, we estimate a mortality rate difference between the exposed group and the unexposed group of -3.95 (95\% CI: -8.46, 0.56; \textit{p}-value = 0.09) per thousand follow-up years using the unadjusted estimator and -4.26 (95\% CI: -8.67, 0.15; \textit{p}-value = 0.06) per thousand follow-up years using CARE.
Using IPW, the estimated mortality rate difference is -5.37 (95\% CI: -16.94, 6.2; \textit{p}-value = 0.36) per thousand follow-up years.
For CARE--IPW the mortality rate difference is -5.08 (95\% CI: -9.46, -0.7; \textit{p}-value = 0.02) per thousand follow-up years.
As in the simulation study, the standard error estimates for the CARE and CARE--IPW are less than those of the unadjusted and IPW.
While IPW had the largest estimated effect size, it also had the largest estimated variance and thereby widest confidence intervals of any estimator.
The estimate made by CARE--IPW was larger than either the unadjusted estimator or CARE and had the smallest variance estimate.

\section{Discussion}

In this paper, we (1) provide non-parametric statistical theory for the covariate-adjusted residuals estimator (CARE) in randomized and observational settings, (2) propose a novel estimator, the covariate-adjusted residuals estimator with inverse probability weighting (CARE--IPW), and (3) support theoretical results with a simulation study and an application to a cluster randomized trial.
Specifically, we prove that CARE is consistent for the average treatment effect (ATE) in randomized studies.
We also prove that CARE is not consistent for the ATE in most observational settings of interest (e.g. when there is an exposure effect).
We develop a new estimator, CARE--IPW, which is consistent for the ATE in observational settings when the propensity scores are consistently estimated.

The simulation study supports our theoretical findings and suggests some advantages to using CARE--IPW rather than CARE or the IPW estimator.
In randomized trials, CARE and CARE--IPW achieved greater precision and statistical power, compared to the other estimators, and  maintained nominal confidence interval coverage.
In observational settings, CARE--IPW is consistent for the statistical parameter $\Psi^{obs}(\mathbb{P})$ when accounting for the confounding covariates in the propensity score model and has greater statistical power and less variability than the IPW estimator.
CARE is biased in observational settings with an exposure effect.

While CARE--IPW improves on CARE and IPW, it is not a ``double robust estimator'', such as targeted maximum likelihood estimation and augmented inverse probability weighting.\cite{robins2000robust, vanderLaan2003, VanderLaan2006, MarkBook}
A double robust estimator is consistent for $\Psi^{obs}(\mathbb{P})$ if either the outcome predictions (which often include the exposure as well as baseline covariates) or the propensity scores is consistently estimated and is the most efficient estimator if both are.
Similar to the IPW estimator, CARE--IPW is consistent for if and only if the propensity score is consistently estimated.
By incorporating predictions of the outcome, CARE--IPW is expected to be more efficient than IPW.

One advantage to using CARE--IPW rather than another method is that researchers do not need to specify the relationship between the exposure and the outcome.
This can be beneficial when there is a complex relationship between the exposure and outcome, such as multiple non-linear interactions with other covariates that augment the strength of the exposure.

As with the IPW estimator, CARE--IPW may have stability issues when estimated propensity scores approach zero or one.\cite{Petersen2010}
This could be resolved in one of a couple ways.
Stabilized weights could be used to scale propensity scores away from zero and one.\cite{Robins2000}
Alternatively, propensity scores could be replaced by incremental propensity scores which relax the positivity assumption by looking at the effect of an intervention when propensity scores are uniformly increased and decreased across all observations.\cite{Kennedy2018}

The findings of this paper suggest that CARE is suitable for use in estimating the average treatment effect in randomized trials, but not in observational settings.
As an alternative to CARE, CARE--IPW has potential for use as an estimator in observational settings; further research is warranted.

\appendix

\section{Proof of Theorem 2.1}
\label{app:care-rct}

Suppose we are in a trial setting.
Let $W^Y$ denote the covariates that are predictive of the outcome, $A$ be a binary indicator of receiving the exposure, and $Y$ be the outcome.
We assume that we have $n$ independent, identically distributed copies of observed data $O=(W^Y,A,Y)$ with some distribution $\mathbb{P}$.
In the following, we assume discrete random variables for simplicity; however, all summations generalize to integrals for continuous random variables.
In a randomized trial, our target statistical estimand is $\Psi^{RCT}(\mathbb{P}) = \mathbb{E}(Y|A=1) - \mathbb{E}(Y|A=0)$.

The expectation of the CARE estimating function (Equation~\ref{eqn:CAREee}) is
\begin{align*}
\mathbb{E} \left[ D(O; \psi ) \right]
=& \mathbb{E} \left[ \left( \frac{\mathbb{I}(A=1)}{\mathbb{P}(A=1)} - \frac{\mathbb{I}(A=0)}{\mathbb{P}(A=0)}\right) \big(Y - \mathbb{E}(Y \mid W^Y) \big) \right] - \psi \\
=& \mathbb{E} \left[ \left( \frac{\mathbb{I}(A=1)}{\mathbb{P}(A=1)} - \frac{\mathbb{I}(A=0)}{\mathbb{P}(A=0)}\right) Y \right]
- \mathbb{E} \left[ \left( \frac{\mathbb{I}(A=1)}{\mathbb{P}(A=1)} - \frac{\mathbb{I}(A=0)}{\mathbb{P}(A=0)}\right) \mathbb{E}(Y \mid W^Y) \right] - \psi
\end{align*}
The first component of the expectation is equal to the target parameter in a randomized trial $\Psi^{RCT}(\mathbb{P})=\psi^{RCT}$:
\begin{align*}
\mathbb{E} \left[ \left( \frac{\mathbb{I}(A=1)}{\mathbb{P}(A=1)} - \frac{\mathbb{I}(A=0)}{\mathbb{P}(A=0)}\right) Y \right] \\
= & \sum_{a,y} \left( \frac{\mathbb{I}(A=1)}{\mathbb{P}(A=1)} - \frac{\mathbb{I}(A=0)}{\mathbb{P}(A=0)}\right) y\mathbb{P}(Y=y \mid A=a)\mathbb{P}(A=a) \\
= & \sum_{y} \frac{\mathbb{P}(A=1)}{\mathbb{P}(A=1)} y \mathbb{P}( Y=y \mid A=1) - \sum_y \frac{\mathbb{P}(A=0)}{\mathbb{P}(A=0)} y \mathbb{P}( Y=y \mid A=0) \\
= & \sum_{y} y \mathbb{P}(Y=y \mid A=1) - \sum_{y} y \mathbb{P}(Y=y \mid A=0) \\
= & \mathbb{E}(Y|A=1) - \mathbb{E}(Y|A=0) \\
= & \psi^{RCT}
\end{align*}

The second component of the expectation is 0:
\begin{align*}
\mathbb{E} & \left[ \left( \frac{\mathbb{I}(A=1)}{\mathbb{P}(A=1)} - \frac{\mathbb{I}(A=0)}{\mathbb{P}(A=0)}\right) \mathbb{E}(Y \mid W^Y) \right] \\
&= \sum_{w^Y,a,y} \left( \frac{\mathbb{I}(A=1)}{\mathbb{P}(A=1)} - \frac{\mathbb{I}(A=0)}{\mathbb{P}(A=0)}\right) y \mathbb{P}( Y=y \mid W^Y=w^Y)\mathbb{P}(A=a \mid W^Y=w^Y)\mathbb{P}(W^Y=w^Y) \\
& = \sum_{w^Y,y} \left[ \frac{\mathbb{P}(A=1 \mid W^Y=w^Y)}{\mathbb{P}(A=1)} - \frac{\mathbb{P}(A=0 \mid W^Y=w^Y)}{\mathbb{P}(A=0)} \right] y \mathbb{P}( Y=y \mid W^Y=w^Y)\mathbb{P}(W^Y=w^Y)
\end{align*}
In a randomized trial, we have $\mathbb{P}(A \mid W^Y)=\mathbb{P}(A)$, and thus the second component is zero.

Thus, when our parameter of interest is $\psi=\psi^{RCT}$, the expectation of the CARE estimating equation \[
\mathbb{E}[D(O\mid \psi^{RCT} )] = \psi^{RCT} - 0 - \psi^{RCT}=0.
\]
This proves that in a trial setting the CARE estimating function is unbiased for the statistical parameter $\psi^{RCT}$, which identifies the average treatment effect because identifiability assumptions hold by design.

\noindent \textbf{Corollary 2.1.1}:
If the predicted outcome $\hat{\mathbb{E}}(Y \mid W^Y)$ is a constant (e.g. 0 or the sample average outcome), CARE reduces to the unadjusted difference in mean outcomes.

\textit{Proof:}
Denote the predicted outcome $\hat{\mathbb{E}}(Y \mid W^Y)$ with a constant $C$, and let $n_1$ and $n_0$ denote the number of treated and control units, respectively. Then we have
\begin{align*}
\hat{\Psi}^{CARE} &= \frac{1}{n} \sum_{i=1}^n \left( \frac{\mathbb{I}(A_i=1)}{\hat{\mathbb{P}}(A=1)} - \frac{\mathbb{I}(A_i=0)}{\hat{\mathbb{P}}(A=0)} \right)(Y_i- C) \\
&= \frac{1}{n} \sum_{i=1}^n \left( \frac{\mathbb{I}(A_i=1)}{\hat{\mathbb{P}}(A=1)} - \frac{\mathbb{I}(A_i=0)}{\hat{\mathbb{P}}(A=0)} \right)Y_i -
\frac{1}{n_1} \sum_{i \in A_i=1} C - \frac{1}{n_0} \sum_{i \in A_i=0} C \\
&= \frac{1}{n} \sum_{i=1}^n \left( \frac{\mathbb{I}(A_i=1)}{\hat{\mathbb{P}}(A=1)} - \frac{\mathbb{I}(A_i=0)}{\hat{\mathbb{P}}(A=0)} \right)Y_i -\frac{n_1}{n_1} C - \frac{n_0}{n_0} C \\
&= \frac{1}{n} \sum_{i=1}^n \left( \frac{\mathbb{I}(A_i=1)}{\hat{\mathbb{P}}(A=1)} - \frac{\mathbb{I}(A_i=0)}{\hat{\mathbb{P}}(A=0)} \right)Y_i.
\end{align*}

\section{Proof of Theorem 2.2}
\label{app:care-obs}

Suppose we are in an observational setting.
Let $W^C$ denote the confounding covariates, $A$ be binary an indicator of receiving the exposure, and $Y$ be the outcome.
We assume that we have $n$ independent, identically distributed copies of observed data $O=(W^C,A,Y)$ with some distribution $\mathbb{P}$.
In the following, we assume discrete random variables for simplicity; however, all summations generalize to integrals for continuous random variables.
In an observational setting, our target statistical estimand is $\Psi^{obs}(\mathbb{P})=\mathbb{E}_{W^C}\big[ \mathbb{E}(Y|A=1, W^C) - \mathbb{E}(Y|A=0,W^C) \big]$.

Using the same steps as in the Proof for Theorem 2.1 (Appendix \ref{app:care-rct}), but replacing the predictive covariates $W^Y$ with the confounding covariates $W^C$, the expectation of the CARE estimating function (Equation~\ref{eqn:CAREee}) in an observational setting is given by
\begin{align*}
\mathbb{E} & \left[ D(O; \psi) \right] \\
=& \mathbb{E} \left[ \left( \frac{\mathbb{I}(A=1)}{\mathbb{P}(A=1)} - \frac{\mathbb{I}(A=0)}{\mathbb{P}(A=0)}\right) Y \right]
- \mathbb{E} \left[ \left( \frac{\mathbb{I}(A=1)}{\mathbb{P}(A=1)} - \frac{\mathbb{I}(A=0)}{\mathbb{P}(A=0)}\right) \mathbb{E}(Y \mid W^C) \right] - \psi \\
= & \psi^{RCT} \\
&- \sum_{w^C,y} \left[ \frac{\mathbb{P}(A=1 \mid W^C=w^C)}{\mathbb{P}(A=1)} - \frac{\mathbb{P}(A=0 \mid W^C=w^c)}{\mathbb{P}(A=0)} \right] y \mathbb{P}( Y=y \mid W^C=w^C)\mathbb{P}(W^C=w^C) \\
& - \psi
\end{align*}
where $\psi^{RCT}=\mathbb{E}(Y|A=1) - \mathbb{E}(Y|A=0)$ and where, due to confounding, $\mathbb{P}(A=a \mid W^C=w^C) \ne \mathbb{P}(A=a)$.
When our parameter of interest is $\psi=\psi^{obs}$, the expectation $\mathbb{E}[D(O; \psi^{obs}]$ is generally not zero.

Under a non-parametric statistical model, we can only guarantee the expectation $\mathbb{E}[D(O; \psi^{obs}]$ is zero under the strong null, where $\psi^{obs}=0$ and $\mathbb{E}(Y \mid A, W^C) = \mathbb{E}(Y \mid W^C)$:
\begin{align*}
\mathbb{E} & \left[ D(O; \psi^{obs}) \right] \\
&= \mathbb{E} \left[ \left( \frac{\mathbb{I}(A=1)}{\mathbb{P}(A=1)} - \frac{\mathbb{I}(A=0)}{\mathbb{P}(A=0)}\right) \big(Y - \mathbb{E}(Y \mid W^C) \big)\right] - \psi^{obs}\\
&=	\mathbb{E} \left[ \left( \frac{\mathbb{P}(A=1 \mid W^C)}{\mathbb{P}(A=1)} - \frac{\mathbb{P}(A=0 \mid W^C)}{\mathbb{P}(A=0)} \right) \left( \mathbb{E}(Y|A,W^C) - \mathbb{E}(Y|W^C) \right) \right] - \psi^{obs}\\
&= 0 \text{ if } \mathbb{E}(Y|A,W^C) = \mathbb{E}(Y|W^C) \text{ and } \psi^{obs}=0
\end{align*}
When the null is false, there might also be some scenarios when $\mathbb{E}\left[ D(O; \psi^{obs}) \right] =0$, but this cannot be proven under a non-parametric statistical model.
Thus, we conclude that the CARE estimating function is generally not an unbiased for $\psi^{obs}$, even if the covariates $W^C$, which are sufficient to control for confounding, are used to predict the outcome.

\section{Proof of Theorem 3.1}
\label{app:care-ipw}

Suppose we are in an observational setting.
Let $W^C$ denote the confounding covariates, $A$ be binary an indicator of receiving the exposure, and $Y$ be the outcome.
We assume that we have $n$ independent, identically distributed copies of observed data $O=(W^C,A,Y)$ with some distribution $\mathbb{P}$.
In the following, we assume discrete random variables for simplicity; however, all summations generalize to integrals for continuous random variables.
In an observational setting, our target statistical estimand is $\Psi^{obs}(\mathbb{P})=\mathbb{E}_{W^C}\big[ \mathbb{E}(Y|A=1, W^C) - \mathbb{E}(Y|A=0,W^C) \big]$.

The expectation of the CARE--IPW estimating function (Equation~\ref{eqn:CAREIPWee}) is
\begin{align*}
\mathbb{E} \left[ D^*(O \mid \psi ) \right]
=& \mathbb{E} \left[ \left( \frac{\mathbb{I}(A=1)}{\mathbb{P}(A=1 \mid W^C)} - \frac{\mathbb{I}(A=0)}{\mathbb{P}(A=0| \mid W^C)}\right) \big(Y - \mathbb{E}(Y \mid W^C) \big) \right] - \psi \\
=& \mathbb{E} \left[ \left( \frac{\mathbb{I}(A=1)}{\mathbb{P}(A=1 \mid W^C)} - \frac{\mathbb{I}(A=0)}{\mathbb{P}(A=0 \mid W^C)}\right) Y \right] \\
& - \mathbb{E} \left[ \left( \frac{\mathbb{I}(A=1)}{\mathbb{P}(A=1\mid W^C)} - \frac{\mathbb{I}(A=0)}{\mathbb{P}(A=0 \mid W^C )}\right) \mathbb{E}(Y \mid W^C) \right] - \psi
\end{align*}
The first component of the expectation is equivalent to the IPW estimand and equal to the target parameter $\Psi^{obs}(\mathbb{P})=\psi^{obs}$:
\begin{align*}
\mathbb{E} &\left[ \left( \frac{\mathbb{I}(A=1)}{\mathbb{P}(A=1 \mid W^C)} - \frac{\mathbb{I}(A=0)}{\mathbb{P}(A=0 \mid W^C)}\right) Y \right] \\
=& \sum_{w^C,a,y} \left( \frac{\mathbb{I}(A=1)}{\mathbb{P}(A=1 \mid W^C=w^C)} - \frac{\mathbb{I}(A=0)}{\mathbb{P}(A=0 \mid W^C)}\right) y \mathbb{P}(Y=y \mid A=a, W^C=w^C)\\
& \times \mathbb{P}(A=a \mid W^C=w^C) \mathbb{P}( W^C=w^C) \\
=& \sum_{w^C,y} \frac{\mathbb{P}(A=1 \mid W^C=w^C)}{\mathbb{P}(A=1 \mid W^C=w^C)} y \mathbb{P}(Y=y \mid A=1, W^C=w^C) \mathbb{P}( W^C=w^C) \\
&- \sum_{w^C,y} \frac{\mathbb{P}(A=0 \mid W^C=w^C)}{\mathbb{P}(A=0 \mid W^C=w^C)} y \mathbb{P}(Y=y \mid A=0, W^C=w^C) \mathbb{P}( W^C=w^C) \\
&= \mathbb{E}_{W^C} \Big [ \mathbb{E}(Y \mid A=1, W^C) - \mathbb{E}(Y \mid A=0, W^C) \Big] \\
&= \psi^{obs}
\end{align*}

The second component of the expectation is 0:
\begin{align*}
\mathbb{E} & \left[ \left( \frac{\mathbb{I}(A=1)}{\mathbb{P}(A=1 \mid W^C)} - \frac{\mathbb{I}(A=0)}{\mathbb{P}(A=0 \mid W^C)}\right) \mathbb{E}(Y \mid W^C) \right] \\
= & \sum_{w^C,a,y} \left( \frac{\mathbb{I}(A=1)}{\mathbb{P}(A=1 \mid W^C=w^C)} - \frac{\mathbb{I}(A=0)}{\mathbb{P}(A=0 \mid W^C=w^C)}\right) y \mathbb{P}( Y=y \mid W^C=w^C) \\
& \times \mathbb{P}(A=a \mid W^C=w^C)\mathbb{P}(W^C=w^C) \\
& = \sum_{w^C,y} \left( \frac{ \mathbb{P}(A=1 \mid W^C=w^C)}{\mathbb{P}(A=1 \mid W^C=w^C)} - \frac{\mathbb{P}(A=0 \mid W^C=w^C)}{\mathbb{P}(A=0 \mid W^C=w^C)} \right) y \mathbb{P}( Y=y \mid W^C=w^C)\mathbb{P}(W^C=w^C) \\
&= 0
\end{align*}

When our parameter of interest is $\psi=\psi^{obs}$, the expectation of the CARE--IPW estimating function is
\[
\mathbb{E}[D^*(O; \psi^{obs} )] = \psi^{obs}- 0 - \psi^{obs} =0\]
This proves that in an observational setting the CARE--IPW estimating function is unbiased for the statistical parameter $\psi^{obs}$, which identifies the average treatment effect under the assumptions outlined in Section~\ref{sec:causal}.
In a randomized trial, we have $\Psi^{obs}(\mathbb{P})=\Psi^{RCT}(\mathbb{P})$, and thus the CARE-IPW estimating function is also unbiased when the exposure mechanism is known.

\section*{Acknowledgements}
We thank Mark van der Laan for his expert advice.

This project was funded by NIH NIAID grant 1R01AI102939 and NIGMS grant R35GM119582.
The findings and conclusions in this manuscript are those of the authors and do not necessarily represent the views of the National Institutes of Health or the National Institute of General Medical Sciences.
The funders had no role in study design, data collection and analysis, decision to present, or preparation of the presentation.

\bibliography{care}

\end{document}

\maketitle

\begin{figure}[H]
\includegraphics[width=0.5\textwidth]{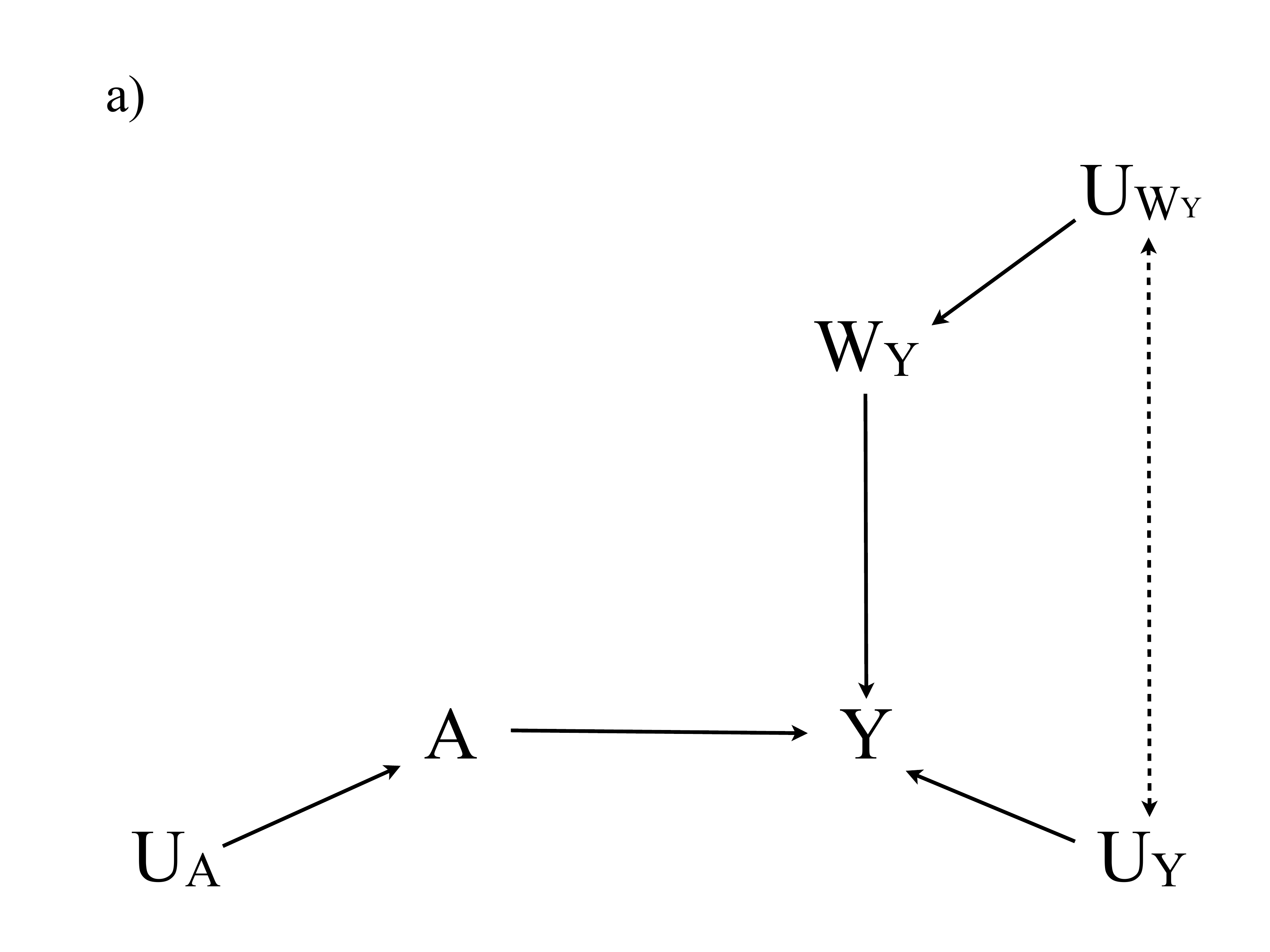}
\includegraphics[width=0.5\textwidth]{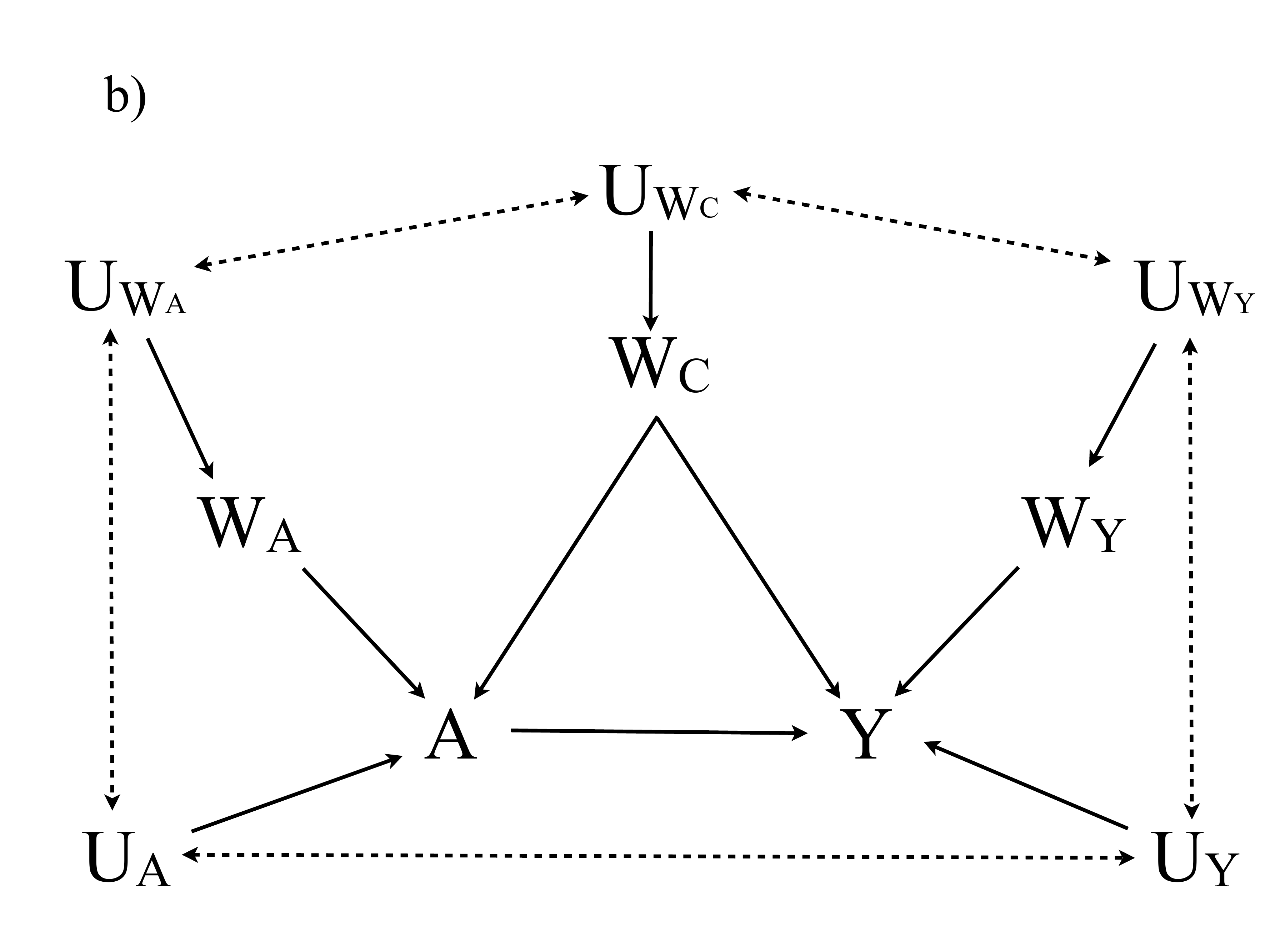}
\caption{
Causal diagrams for randomized trials \textbf{(a)} and observational studies \textbf{(b)} including measured and unmeasured covariates.
These diagrams give us a visual representation of the relationships between the variables in a causal model.
Arrows are drawn from a cause to an effect; dashed double-sided arrows indicate an unknown or unmeasured relationship.
In a randomized setting, the exposure of interest ($A$) is independent of all other variables and the outcome of interest ($Y$) is influenced by both $A$ and a set of other covariates ($W^Y$).
Randomization also guarantees that the unmeasured factors influencing $A$ ($U_A$) are independent of the unmeasured factors influencing $W^Y$ ($U_{W^Y}$) and $Y$ ($U_Y$).
In an observational setting, $A$ is no longer randomized, but instead influenced by other covariates.
Some of these covariates ($W^C$) also influence $Y$, thus confounding the relationship between $A$ and $Y$.
Other covariates ($W^A$) only influence $A$ and not $Y$.
Without randomization any of the unmeasured covariates may have a relationship with any of the other unmeasured covariates, as indicated by the dashed arrows around the perimeter of the diagram.
}
\label{fig:dag-u}
\end{figure}